\documentclass[11pt]{article}
 
\usepackage{amsmath,amsthm}
\usepackage{amsthm}
\usepackage{amssymb}
\usepackage{amsfonts} 

\usepackage[latin1]{inputenc}

\usepackage{graphicx}

\newtheorem{theorem}{Theorem}

\def\be{\begin{equation}}
\def\ee{\end{equation}}
\def\bea{\begin{eqnarray}}
\def\eea{\end{eqnarray}}

\newcommand{\sect}[1]{\setcounter{equation}{0}\section{#1}}
\newcommand{\subsect}[1]{\subsection{#1}}

\newcommand{\bq}{\mathbf{q}}
\newcommand{\bp}{\mathbf{p}}

\newcommand{\te}{\theta}

\newcommand{\hbq}{\hat{\mathbf{q}}}
\newcommand{\hbp}{\hat{\mathbf{p}}}
\newcommand{\hH}{\hat{\cal{H}}}
\newcommand{\hC}{\hat{C}}
\newcommand{\hI}{\hat{I}}

\newcommand{\hq}{\hat{q}}
\newcommand{\hp}{\hat{p}}

\newcommand{\hr}{\hat{r}}
\newcommand{\hte}{\hat{\te}}

\def\RR{\mathbb{R}}

\newcommand{\cM}{{\mathcal M}}
\def\dd{{\rm d}}
\newcommand\Om\Omega
\newcommand{\bL}{\mathbf{L}}

\newcommand{\hbL}{\hat{\mathbf{L}}}

\newcommand{\De}{\Delta}

\newcommand{\pd}{\partial}
\newcommand\minus\backslash

\newcommand{\om}{\omega}

\newcommand{\LB}{{\rm TLB}}

\def\la{{\lambda}}

\def\rmi{{\rm i}}
\def\rme{{\rm e}}
\def\ff{f}

\def\hh{v}

\def\ur{u}

\def\pot{{\cal U}_2}
\def\vpot{{\cal V}_2}

\def\hamG{\hat{\cal H}_{\rm TLB}}

\def\1{\'{\i}}

\def\k{{\kappa}}

\def\>#1{{\mathbf#1}}

\def\uu{{\cal U}_{1}}
\def\vv{{\cal V}_{1}}

\def\phiLB{\Phi_{\rm TLB}}

\def\phiPDM{\Phi_{\rm TPDM}}

\def\phiS{\Phi}

\def\omm{\Omega}

\def\bb{\beta}

\def\arc{{\rm arc}\!}
\def\NN{\mathbb N}
\def\cY{\mathcal{Y}}
\def\SS{\mathbb S}
\DeclareMathOperator\id{id}

\DeclareMathOperator\spec{spec}

\begin{document}

 \
 \smallskip

  \noindent
{\Large{\bf{Quantum mechanics on spaces of nonconstant curvature:  \\[6pt] the oscillator problem and superintegrability}}}

\bigskip

\bigskip

\begin{center}
{\large  Ángel Ballesteros$^a$,   Alberto Enciso$^b$,  Francisco J. Herranz$^a$,\\[4pt] Orlando Ragnisco$^c$ and Danilo Riglioni$^c$}
\end{center}

\noindent
{$^a$ Departamento de F\1sica,  Universidad de Burgos,
09001 Burgos, Spain\\ ~~E-mail: angelb@ubu.es \quad fjherranz@ubu.es\\[10pt]
}
$^b$ Departamento de F\1sica Teórica II,   Universidad Complutense,   28040 Madrid,
Spain\\ ~~E-mail: aenciso@fis.ucm.es\\[10pt]
$^c$ Dipartimento di Fisica,   Università di Roma Tre and Istituto Nazionale di
Fisica Nucleare sezione di Roma Tre,  Via Vasca Navale 84,  00146 Roma, Italy  \\
~~E-mail: ragnisco@fis.uniroma3.it  \quad riglioni@fis.uniroma3.it    \\[10pt]

\medskip
\medskip

\begin{abstract}
\noindent 
The full spectrum and eigenfunctions of the quantum version of a nonlinear oscillator defined on an $N$-dimensional space with nonconstant curvature are rigorously found. Since the underlying curved space generates a position-dependent kinetic energy, three different quantization prescriptions are worked out by imposing that the maximal superintegrability of the system has to be preserved after quantization. The relationships among these three Schrödinger problems are described in detail through appropriate similarity transformations. These three approaches are used to illustrate different features of the quantization problem on $N$-dimensional curved spaces or, alternatively, of position-dependent mass quantum Hamiltonians. This quantum oscillator is, to the best of our knowledge, the first example of a maximally superintegrable quantum system on an $N$-dimensional space with nonconstant curvature.

 \end{abstract}

\bigskip\bigskip\bigskip\bigskip



\newpage

\sect{Introduction}

 This paper is devoted to the study of the quantum mechanical version of $N$-dimensional ($N$D) classical Hamiltonian systems of the type 
$$
{\cal H}(\bq,\bp)={\cal T}(\bq,\bp)+{\cal U}(\bq)=
\frac{\bp^2}{2\, {\cal M}(\bq)}+{\cal U}(\bq) ,
 $$
where $\bq,\bp\in\RR^N$ are conjugate coordinates and momenta with canonical Poisson  bracket $\{q_i,p_j\}=\delta_{ij}$. The physical interpretation of these systems is two-fold. On one hand, as the Hamiltonian describing the motion of a particle on the $N$D curved space defined by the (conformally flat) metric $\dd s^2= {\cal M}(\bq)\,\dd \bq^2$ and under the action of the potential ${\cal U}(\bq)$. On the other, as position-dependent mass systems on the $N$D Euclidean space.

Evidently, the crucial point for the definition of the corresponding Schrödinger problem is the consistent (under certain given criteria) definition of the quantum kinetic energy term, ${\cal T}(\bq,\bp)\to \hat{\cal T}(\hat\bq,\hat\bp)$,  since an obvious ordering ambiguity appears when the position and momenta operators are considered.  In this paper we shall deal with a specific example:
the motion of a particle on the so-called $N$D Darboux III space~\cite{PLB, annals} given by ${\cal M}(\bq)=1+\la\, \bq^2$ (with $\la>0$), and whose potential will be the intrinsic oscillator on such space defined by 
$$
{\cal U}(\bq)=\frac{ \om^2 \bq^2}{2(1+\la \bq^2)}.
$$

This choice is motivated by the fact that this system is the only known example of a maximally superintegrable classical Hamiltonian on an $N$D space with nonconstant curvature~\cite{PhysD}, and it can be interpreted as a $\la$-deformation of the flat isotropic oscillator, which is recovered in the $\la\to 0$ limit. In fact, it is well-known that the quantum superintegrability of the $N$D flat isotropic oscillator is useful in order to obtain its exact solution by making use of the superabundance of quantum integrals of the motion~\cite{Fradkin}. 
Therefore, it seems natural to quantize the Darboux III oscillator in such a way that the maximal superintegrability of the system will be manifestly preserved under quantization; moreover, presumably in this way  this new $N$D nonlinear oscillator could be fully solved by mimicking the standard procedure for the Euclidean oscillator. 

In the sequel we will show that  this is indeed the case. In fact, the superintegrability constraint will be useful in order to analyse in detail several possible quantization prescriptions amenable for $N$D spaces with nonconstant curvature, namely:
 
 \begin{itemize}
 
 \item The so-called~\cite{uwano} direct ``Schrödinger" quantization
 \be
 {\hH}= 
 \frac{1}{2(1+\la \hbq^2)}\, \hbp^2+ \  \frac{ \om^2 \hbq^2}{2(1+\la \hbq^2)} =\frac{1}{2(1+\la \bq^2)}\big(-\hbar^2\De+\om^2\bq^2\big),
 \label{ca0}
 \ee
 which was the one used in~\cite{FTC} for this system, since it preserves the maximal superintegrability in a straightforward way due to the immediate quantum transcription of the $(2 N-1)$ classical integrals of the motion. This property leads to a maximal degeneracy of the spectrum, which is exactly the same as in the quantum $N$D flat oscillator.
 
 \item The ``Laplace--Beltrami" (LB) quantization, which makes use of the usual LB operator on curved spaces:
\be
\hat{\cal H}_{\rm LB}=-\frac {\hbar^2} 2 \Delta_{\rm LB}+ \frac{\omega^2 \mathbf{q}^2}{2(1+\lambda \mathbf{q}^2)} \qquad\mbox{where} \quad 
\Delta_{\rm LB}=\sum_{i,j=1}^N \frac 1{\sqrt{g}}\partial_i\sqrt{g} g^{ij}\partial_j .
\label{LB0}
\ee
 However, we shall show that this LB Hamiltonian cannot be transformed into (\ref{ca0}) through a similarity transformation unless we include an additional quantum potential which is proportional  to the scalar curvature of the underlying space. This similarity transformation guarantees that the spectra of (\ref{LB0}) and (\ref{ca0}) coincide, and provides the explicit form of the full set of quantum integrals for $\hat{\cal H}_{\rm LB}$. We stress that such a kind of  quantum ``geometric" potential is well-known in the literature (both in scalar field theories in General Relativity as well as in the context of the quantization problem on generic Riemannian manifolds~\cite{Wa84, Landsman, Liu}), and it is tantamount to replace the LB operator by   the so-called ``conformal Laplacian"~\cite{Baer}. We have to mention here that the connection between LB operators and scalar curvatures associated with two different, possibly conformally flat, Riemannian manifolds has been firstly pointed out in a pioneering paper by Paneitz in 1983~\cite{Paneitz}.  However, the neat connection of the latter with quantum superintegrability properties is here stated for the first time and we think that this result opens the path for a novel algebraic approach to the subject. 
 
 \item Finally, a ``position-dependent mass" (PDM) quantization, which is essential in many condensed matter problems (see for instance~\cite{Bastard, qDWW, Roos, mass2, Quesnea, Quesnec}). Here we perform it by considering the symmetric prescription proposed in~\cite{mass2}, namely:
\be
  \hat{\cal H}_{\rm PDM}(\hat\bq,\hat\bp)=\frac { 1} 2\, \hbp\cdot \frac{1}{(1+\la \hbq^2)} \hbp
  + \  \frac{ \om^2 \hbq^2}{2(1+\la \hbq^2)} =-\frac {\hbar^2} 2 \nabla\cdot  \frac{1}{(1+\la \bq^2)} \nabla  + \frac{\omega^2 \mathbf{q}^2}{2(1+\lambda \mathbf{q}^2)}.
\label{oa0}
\ee
In this case we  again  find that, in order to get a similarity transformation leading to (\ref{ca0}), another additional quantum potential has to be added to (\ref{oa0}). In this way the spectrum is preserved and the full set of quantum integrals of the motion is explicitly obtained.
 
 \end{itemize}
 
 We stress that although the three previous quantum Hamiltonians have different explicit expressions and, to some extent, interpretations, 
 all of them are related through similarity transformations provided the additional quantum potential terms are considered. This, in turn, means that  they share a common energy spectrum, but they have different wave functions. Moreover, one of the main objectives of this comparative analysis is to point out some generic ({\em i.e.} potential independent) features of the quantization problem on curved spaces, such as the relevance of the dimension $N$ of the underlying manifold (the $N=2$ case will be distinguished) as well as the particular properties of the nonconstant curvature cases.

The paper is organized as follows. In the next Section, the classical Darboux III oscillator is revisited in order to provide the necessary classical background on this system, including the explicit description of its maximal superintegrability in terms of a curved Fradkin tensor. Section 3 is devoted to review the geometry of the underlying space, emphasizing the role of its nonconstant curvature and introducing a classical radial effective potential that will be useful in the quantum context. In Section 4 the three superintegrable quantizations of the Darboux III oscillator are obtained, and the similarity transformations among them are fully described. Section 5 provides the three associated radial Schrödinger equations. The spectrum and eigenfunctions of the system are rigorously obtained in Section 6, thus completing the preliminary results given in~\cite{FTC}. Finally, Section 7 includes several remarks and open problems.

 
\sect{The Darboux III oscillator}

 The $N$D classical Hamiltonian system given by
 \be
H(\bq,\bp)=\frac{\bp^2+\om^2\bq^2}{\k+\bq^2} ,
 \label{aa}
 \ee
 with   real parameters $\k>0$ and $\om\geq 0$,  was proven in~\cite{PhysD} to be maximally superintegrable (MS), since this Hamiltonian is endowed with the maximum possible number of  $2N-1$ functionally independent constants of motion. Hereafter we shall consider the equivalent Hamiltonian $\cal H$ defined by ${\cal H}=  \k H/ 2$   with real parameter $\la=1/\k>0$:
\be
{\cal H}(\bq,\bp)={\cal T}(\bq,\bp)+{\cal U}(\bq)=
\frac{\bp^2}{2(1+\la \bq^2)}+\frac{ \om^2 \bq^2}{2(1+\la \bq^2)}  .
 \label{ac}
\ee

The kinetic energy $ {\cal T}(\bq,\bp)$ can be interpreted as the one generating the geodesic motion of a particle with unit mass  on a conformally flat space with metric and (nonconstant) scalar curvature  given by
 \be
 \dd s^2= (1+\la \bq^2)\dd \bq^2 , \qquad  R(\bq)=-\la\,\frac{(N-1)\bigl( 2N+3(N-2)\la \bq^2\bigr)}{(1+\la \bq^2)^3} .
 \label{ad}
 \ee
 In fact, such a curved space is the $N$D spherically symmetric generalization of the Darboux surface of type III~\cite{Ko72,KKMW03}, which was constructed in~\cite{PLB, annals}.  On the other hand,   the central potential $ {\cal U}$ was proven in~\cite{annals,PhysD} to be an ``intrinsic" oscillator potential on that Darboux space.

 Moreover, in spite of the very naive appearance of the classical Hamiltonian  (\ref{ac}), which is nothing but the $N$D isotropic harmonic oscillator system divided by its oscillator potential (plus a rather {\em relevant} constant here scaled to 1),  it is worth mentioning that this system can also be considered in three other different (but related) frameworks:
 
 \begin{itemize}
 
 \item  For $N=3$,  $\cal H$  arises as a particular   case of the so call  multifold (or $\nu$-fold) Kepler 3D  Hamiltonians constructed in~\cite{uwano,IK95} as generalizations of  the MIC--Kepler and   Taub-NUT  systems~\cite{mic1,mic2,mic3,mic4,mic5,mic6,mic7,mic8,mic9}. In the notation of~\cite{IK95}, 
   $\cal H$ can be recovered by setting $\nu=1/2$, $a=1$ and $b=\la$. Notice that from our approach, the proper Kepler--Coulomb potential on the Darboux space (\ref{ad}) would be~\cite{annals}
 $$
 {\cal U}_{\rm KC}=\alpha\frac{\sqrt{1+\la \bq^2}}{|\bq|},\qquad |\bq|=\sqrt{\bq^2},\qquad \alpha \in\mathbb R,
 $$
which is related  with the curved oscillator potential through ${\cal U}\propto
 {\cal U}_{\rm KC}^{-2}$.
 
  \item  Again for $N=3$,  the potential $\cal U$  can be obtained from the temporal part of  the family of the so called Bertrand metrics~\cite{Perlick, Bertrand} on  $(3+1)$D Lorentzian spacetimes with nonconstant curvature. Such $(3+1)$D free systems   possess stable circular orbits and all of their
bounded trajectories are   periodic, so these are the natural generalization of  the classical Bertrand's theorem~\cite{Bertrand2} to spaces of nonconstant curvature. We recall that the MS property for of all  3D Bertrand Hamiltonians, which come from the  $(3+1)$D Bertrand metrics, was recently proven in~\cite{commun} (see also~\cite{commun2}).
 
 \item As we have already mentioned, 
 $\cal H$  can alternatively be interpreted as describing a 
 PDM system in which the conformal factor of the metric (\ref{ad})  is identified with the mass function, which in this case is parabolic: ${\cal M}(\bq)=1+\la \bq^2$.

 \end{itemize}

We point out that we have chosen to deal with (\ref{ac}) instead of (\ref{aa}) because in this way all the expressions that we shall present throughout the paper  will have  a smooth  and well defined   limit $\la\to 0$ that leads to the well-known results concerning the (flat) $N$D isotropic harmonic oscillator with frequency $\om$. In particular, the limit  $\la\to 0$ of (\ref{ac}) and (\ref{ad})  yields
 \be
 {\cal H}_0=\frac 12 \bp^2+\frac 12 \om^2\bq^2 , \qquad  \dd s^2= \dd \bq^2 ,\qquad R=0.
 \label{ae}
 \ee


\subsect{Maximal superintegrability from a curved Fradkin tensor}

The fact that  $\cal H$ is a MS Hamiltonian can be explicitly demonstrated as follows~\cite{PhysD,IJTP}.

\begin{theorem}
  (i) The Hamiltonian ${\cal H}$  (\ref{ac}) is endowed with the following constants of motion.

\noindent
$\bullet$ $(2N-3)$  angular momentum integrals:
\be
  C^{(m)}=\!\! \sum_{1\leq i<j\leq m} \!\!\!\! (q_ip_j-q_jp_i)^2 , \qquad 
 C_{(m)}=\!\!\! \sum_{N-m<i<j\leq N}\!\!\!\!\!\!  (q_ip_j-q_jp_i)^2 ,   \label{af}
 \ee
 where $m=2,\dots,N$ and $C^{(N)}=C_{(N)}$.
 
 \noindent
$\bullet$ $N^2$ integrals  which form the ND curved Fradkin tensor:
 \be
 I_{ij}=p_ip_j-\bigl(2\la  {\cal H}(\bq,\bp)-\om^2\bigr) q_iq_j , 
\label{ag}
\ee
where $ i,j=1,\dots,N$ and 
such that  ${\cal H}=\frac 12 \sum_{i=1}^N I_{ii}$.  

\noindent
(ii) Each of the three  sets $\{{\cal H},C^{(m)}\}$,  
$\{{\cal H},C_{(m)}\}$ ($m=2,\dots,N$) and   $\{I_{ii}\}$ ($i=1,\dots,N$) is  formed by $N$ functionally independent functions  in involution.

\noindent
(iii) The set $\{ {\cal H},C^{(m)}, C_{(m)},  I_{ii} \}$ for $m=2,\dots,N$ with a fixed index $i$    is  constituted  by $2N-1$ functionally independent functions. 
\end{theorem}

Notice that the first set of $2N-3$ integrals (\ref{af}) is the same for {\em any} central potential on any spherically symmetric space~\cite{annals} since it is provided by an underlying $\mathfrak{sl}(2,\RR)$ coalgebra symmetry (also by an $\mathfrak{so}(N)$-symmetry), while the second one (\ref{ag}) comes from the specific  oscillator potential that we consider here. The latter, in fact,    correspond to a  curved analog of   the Fradkin tensor of integrals of motion~\cite{Fradkin} for the isotropic harmonic oscillator. We also recall that the Hamiltonian (\ref{ac}) together with   both sets of integrals   of   (\ref{af}) and (\ref{ag}) can alternatively   be obtained~\cite{IJTP} from   the  {free Euclidean  motion} by means of a Stäckel transform or coupling constant metamorphosis (see~\cite{Hietarinta, Stackel2, Stackel4, Kalnins1} and references therein).

Thus, in general, the latter integrals (\ref{ag})  do not exist for a generic central potential so that, in principle,  the MS property is not ensured at all.     From this viewpoint the $N$D nonlinear oscillator Hamiltonian $\cal H$ (\ref{ac}) can be regarded as the ``closest  neighbour" of nonconstant curvature to the isotropic harmonic oscillator system (\ref{ae})  (with $\la=0$) as both share the same  MS property.   In fact, the real parameter $\la$ behaves as a ``deformation"  parameter  governing the nonlinear behaviour of $\cal H$, and this parameter is deeply related to the  variable curvature of the underlying Darboux  space.

 
 \subsection{Expressions in terms of hyperspherical coordinates in   phase space}

 The above results can also be expressed in terms of hyperspherical coordinates $r,\te_j$, and canonical   momenta $p_r,p_{\te_j}$,   $(j=1,\dots,N-1)$.
 The $N$ hyperspherical coordinates are formed by a radial-type one $r=|\bq|\in \RR^+$  and   $N-1$   angles $\theta_j$  such that  
  $\theta_k\in[0,2\pi)$ for $k<N-1$ and  $\theta_{N-1} \in[0,\pi)$. These are  defined by
 \begin{equation}
q_j=r \cos\te_{j}     \prod_{k=1}^{j-1}\sin\te_k ,\quad 1\leq j<N,\qquad 
q_N =r \prod_{k=1}^{N-1}\sin\te_k ,
\label{ba}
\end{equation}
where hereafter any product $\prod_{l}^m$ such that $l>m$ is assumed to be equal to 1.
The metric (\ref{ad}) now adopts the form
\begin{equation}
\dd s^2= (1+\la r^2)(\dd r^2+r^2\dd\Om^2),
\label{bb}
\end{equation}
where   $\dd\Om^2$  is the   metric on the unit $(N-1)$D sphere $\SS^{N-1}$
$$
\dd\Om^2=\sum_{j=1}^{N-1}\dd\te_j^2\prod_{k=1}^{j-1}\sin^2\te_k .
\label{bc}
$$
 The relations between  $\>p$ and   $p_r,p_{\te_j}$  read $(1\le j<N)$~\cite{annals}:
\bea
&& p_j=\prod_{k=1}^{j-1}\sin\te_k\cos\te_j\,p_r+\frac{\cos\te_j}{r}\sum_{l=1}^{j-1} \frac{\prod_{k=l+1}^{j-1}\sin\te_k }{\prod_{m=1}^{l-1}\sin\te_m} \cos\te_l\,p_{\te_l}-\frac{\sin\te_j}{r\prod_{k=1}^{j-1}\sin\te_k}\,p_{\te_j} , \nonumber\\
&&p_N=\prod_{k=1}^{N-1}\sin\te_k \,p_r+\frac 1 r\sum_{l=1}^{N-1}\frac{\prod_{k=l+1}^{N-1}\sin\te_k}{\prod_{m=1}^{l-1}\sin\te_m}\cos\te_l \, p_{\te_l} ,
\label{bd}
\eea
 where from now on any sum   $\sum_{l}^m$ such that $l>m$ is assumed to be zero. From (\ref{bd}) we obtain that
  \begin{equation}
\bp^2= p_r^2+r^{-2}\bL^2,
\label{be}
\end{equation}
where $\bL^2$ is the total   angular momentum   given by
\begin{equation}
\bL^2=\sum_{j=1}^{N-1}p_{\te_j}^2\prod_{k=1}^{j-1}\frac{1}{\sin^{2}\te_k}.
\label{bf}
\end{equation}
 By introducing (\ref{ba}) and (\ref{bd}) in    the Hamiltonian (\ref{ac}) we find
   \be
 {\cal H}(r,p_r)= 
 \frac{p_r^2+r^{-2}\bL^2 }{2(1+\la r^2)} +  \frac{ \om^2 r^2}{2(1+\la r^2)} ={\cal T}(r,p_r)+{\cal U}(r) .
 \label{bg}
 \ee
The integrals of motion $C_{(m)}$  (\ref{af})   adopt a compact form (the remaining $C^{(m)}$  and $I_{ij}$  have more cumbersome expressions):
$$
C_{(m)}=\sum_{j=N-m+1}^{N-1}p_{\te_j}^2\prod_{k=N-m+1}^{j-1}\frac 1{\sin^{2}\te_k  },\quad m=2,\dots,N;
\label{bh}
$$
and $C_{(N)} =\bL^2$, which is just the second-order Casimir of the $\mathfrak{so}(N)$-symmetry algebra of a central potential.

Furthermore, the
complete integrability determined by the set of $N$ functions $\{{\cal H},C_{(m)}\}$
$(m=2,\dots, N)$ leads   to a separable set of $N$ equations, since each of them depends on a unique pair of canonical variables. These are the $N-1$ angular  equations
\bea
&& C_{(2)}(\te_{N-1},p_{\te_{N-1}})=p^2_{\te_{N-1}},\nonumber\\[2pt]
&& C_{(k)}(\te_{N-k+1},p_{\te_{N-k+1}})=p^2_{\te_{N-k+1}} +
\frac{C_{(k-1)} }{\sin^{2}\te_{N-k+1}},  \qquad k=3,\dots,N-1 , \nonumber \\[2pt]
&& C_{(N)}(\te_{1},p_{\te_{1}})=p^2_{\te_{1}} +
\frac{C_{(N-1)} }{\sin^{2}\te_1}\equiv  \bL^2 ,
\label{bi}
\eea
together with the single radial equation corresponding to the 1D Hamiltonian  (\ref{bg}).

 
 \sect{The Darboux space and the classical  effective potential}
\label{S.S2}

\begin{figure}
\begin{center}
\includegraphics[height=6cm]{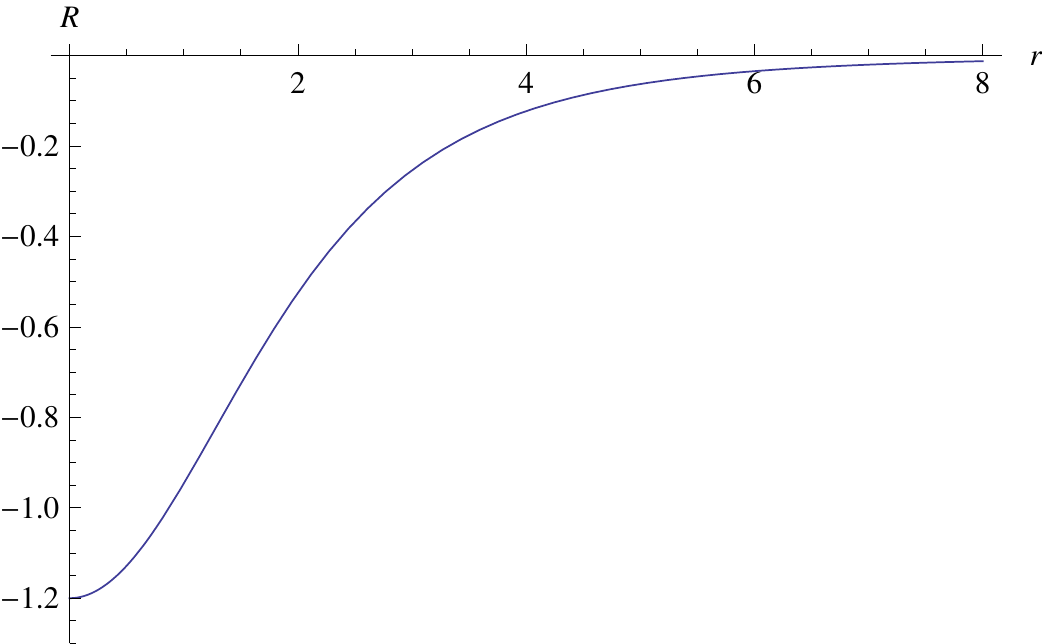}
\caption{Shape of the scalar curvature  (\ref{ad}) of the Darboux space where $r=|\bq|$  for $N=3$ and  $\la=0.1$. The minimum  is always located at the origin, and its value in this case is $R(0)=-1.2$.
 \label{figure1}}
\end{center}
\end{figure}

 The underlying manifold of the classical Hamiltonian (\ref{ac})  is the $N$D  Darboux space with metric (\ref{ad}), whose kinetic energy corresponds to the geodesic motion on the complete Riemannian manifold $\cM^N=(\RR^N,g)$ with
\be
g_{ij}:=(1+\la\bq^2)\,\delta_{ij} ,
\label{metric1}
\ee
and provided that $\la>0$.
The scalar curvature $R(r)\equiv R(|\bq|)$ (\ref{ad}) coming from this metric is always a  {\em negative increasing} function such that
 $\lim_{r\to \infty}R=0$ and  it has a minimum at the origin $$R(0)=-2\la N(N-1),$$ which is exactly the scalar curvature of the $N$D {\em hyperbolic space} with negative constant sectional curvature equal to $-2\la$ (see figure~\ref{figure1}). Recall that the four Darboux surfaces  are the only 2D spaces of nonconstant curvature whose geodesic motion is (quadratically) MS, therefore they are the ``closest" ones to the classical Riemannian spaces of   constant curvature~\cite{Ko72, KKMW03}.

\begin{figure}
\begin{center}
\includegraphics[height=7cm]{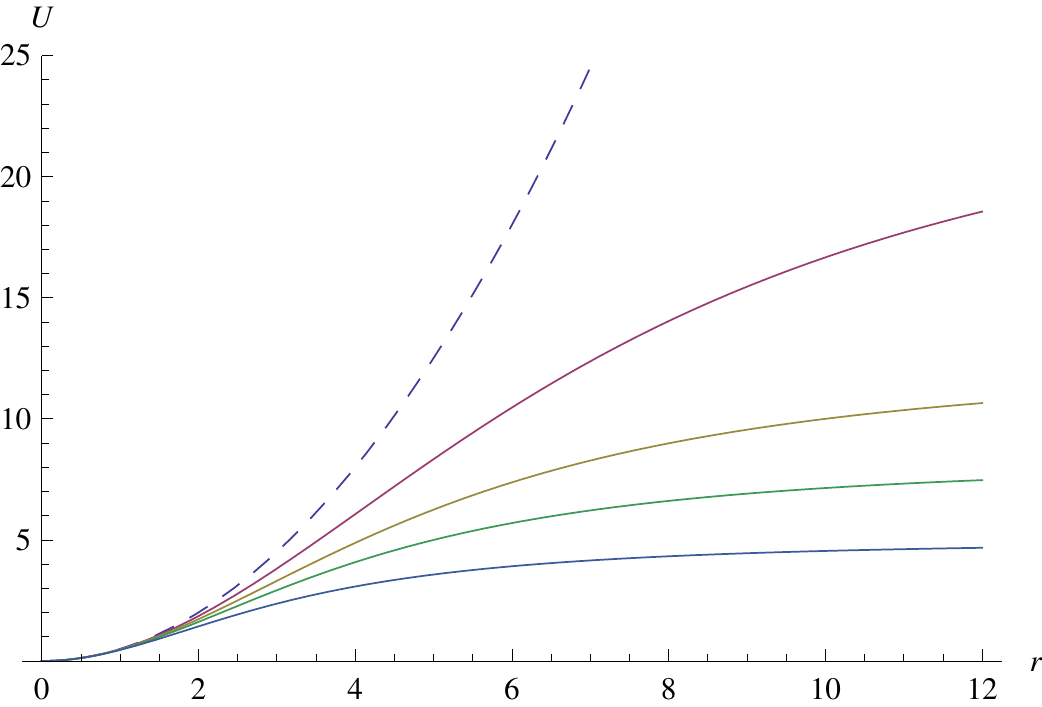}
\caption{The {nonlinear   oscillator potential} (\ref{pota}) with $\omega=1$  for $\la=\{0,\, 0.02,\, 0.04,\, 0.06,\, 0.1\}$ starting from the upper   dashed  line corresponding to the isotropic harmonic oscillator  with $\la=0$. The limit $r\to \infty$ gives   $\{ +\infty, \,25,\,12.5,\, 8.33,\, 5\}$, respectively.
 \label{figure2}}
\end{center}
\end{figure}

As far as the nonlinear radial oscillator potential ${\cal U}(r)$ (\ref{bg})  is concerned, we find    that it is   a  {\em positive increasing} function of $r$,   such that 
\be
{\cal U}(r)=  \frac{ \om^2 r^2}{2(1+\la r^2)}     ,\qquad {\cal U}(0)=0 ,\qquad \lim_{r\to \infty}{\cal U}(r)=\frac{\om^2}{2\la} .
\label{pota}
\ee
 This potential is  shown in figure  \ref{figure2}   for several values of $\la$. Consequently, in contrast with the (Euclidean) isotropic harmonic oscillator, ${\cal U}(r)$ yields a nonlinear behavior
governed  by $\la$, which means that the oscillator potential has the asymptotic maximum 
$\om^2/(2\la)$.


\begin{figure}
\includegraphics[height=7cm,width=10cm]{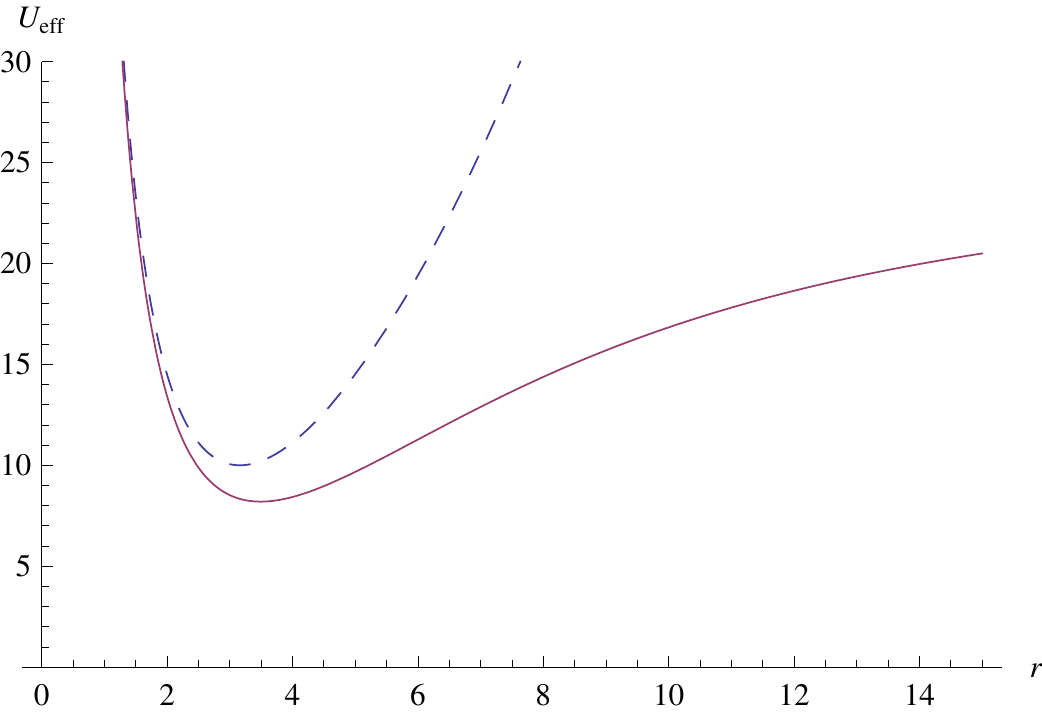}
\caption{The  {classical effective  nonlinear   oscillator potential}  (\ref{bbgg}) for    $\la=0.02$, $c_N=100$ and $\om=1$. The minimum of the potential is located at $r_{\rm min}=3.49$ with ${\cal U}_{\rm eff}(r_{\rm min})=8.2$ and ${\cal U}_{\rm eff}(\infty)=25$.  The dashed  line   corresponds to the effective potential of the    harmonic oscillator  with $\la=0$ with minimum ${\cal U}_{\rm eff}(r_{\rm min})=10$ at  $r_{\rm min}=3.16$. 
 \label{figure3}}
\end{figure}


Nevertheless, since the underlying manifold  $\cM^N$  is not flat, the interplay between the oscillator potential ${\cal U}(r)$ and the kinetic energy term is rather subtle. For this reason, the   complete classical system   can be better understood by introducing a classical effective potential. This can be achieved by applying the 1D  canonical transformation defined by
\be
P(r,p_r)=\frac{p_r}{\sqrt{1+\la r^2}} ,\qquad Q(r)=\frac 12 r\sqrt{1+\la r^2}+\frac{\arc\sinh(\sqrt{\la}r)}{2\sqrt{\la}},
\label{bbhh}
\ee
(where the new canonical variables fulfill $ \{Q,P\}=1$), to the   radial Hamiltonian (\ref{bg}). Notice that
$Q(r)$  has a unique (continuously  differentiable) inverse $r(Q)$, on the whole positive semiline, that is, both $r,Q\in [0,\infty)$ and    
 $\dd Q (r) = \sqrt{1+\la r^2}\dd r$. In this way, we obtain that
  \be
 {\cal H}(Q,P)= \frac 12 P^2+ {\cal U}_{\rm eff}(Q ),\qquad  {\cal U}_{\rm eff}(Q(r)  )= \frac{ c_N }{2(1+\la r^2)r^2} + \ \frac{ \om^2 r^2}{2(1+\la r^2)},
 \label{bbgg}
 \ee
 where the constant $c_N\ge 0$ is the value of the integral of motion corresponding to the square of the total angular momentum $C_{(N)}\equiv \bL^2 $ (\ref{bi}).
Hence the classical system can be described as a particle on  a 1D flat space under  the effective potential $ {\cal U}_{\rm eff}(Q (r))$, which is represented in figure~\ref{figure3}.

 The analysis of ${\cal U}_{\rm eff}$ shows that  this is always positive and it has a minimum located at $r_{\rm min}$ such that
\be
r^2_{\rm min}=\frac{\la c_N+\sqrt{\la^2 c_N^2+\om^2 c_N}}{\om^2},\qquad    {\cal U}_{\rm eff}(Q (r_{\min}))=-\la c_N+\sqrt{\la^2 c_N^2+\om^2 c_N} .
\label{ya}
\ee
Therefore,   $r_{\rm min}$ and $  {\cal U}_{\rm eff}(Q (r_{\min}))$  are, in this order, greater and smaller than those corresponding to the isotropic harmonic oscillator:
\be
 \la=0:\quad r^2_{\rm min}=\frac{ \sqrt{  c_N}}{\om},\qquad    {\cal U}_{\rm eff}(Q (r_{\min}))= \om \sqrt{  c_N} .
\label{yb}
\ee
 Moreover  ${\cal U}_{\rm eff}$ has  two
representative limits:
\be
\lim_{r\to 0} {\cal U}_{\rm eff}(Q (r))=+\infty,\qquad \lim_{r\to \infty} {\cal U}_{\rm eff}(Q (r))=\frac{\om^2}{2\la} ,
\label{bz}
\ee
the  latter being the same of (\ref{pota}).  Thus, this  effective potential is hydrogen-like and one should expect that its quantum counterpart should have both bounded and unbounded states.
The rest of the paper is devoted to solve such a quantum problem in full detail.


 \sect{Superintegrable  quantizations of the Darboux III  oscillator}

  Let us consider 
the quantum position and momenta operators, $\hbq$, $\hbp$, with Lie brackets and   differential representation given by
\be
[\hq_i,\hp_j]=\rmi \hbar \delta_{ij},\qquad \hq_i=q_i,\qquad \hp_i=-\rmi  \hbar \frac{\partial}{\partial q_i}.
\label{ka}
\ee
Hereafter we will use the standard notation 
$$
\nabla=\left(\frac\pd{\pd q_1},\dots,\frac\pd{\pd q_N}\right),\qquad \De=\nabla^2=\frac{\pd^2}{\pd^2 q_1}+\cdots +\frac{\pd^2}{\pd^2 q_N}.
$$


 \subsection{The Schrödinger quantization}

 The so-called ``direct" quantization approach  can be summarized in the following way~\cite{FTC} (this result is worth to be compared with Theorem 1, but taking into account that hereafter the order of the terms becomes crucial).

  \begin{theorem}
    Let   $\hH$ be the quantum Hamiltonian given by
 \be
 {\hH}= 
 \frac{1}{2(1+\la \hbq^2)}\, \hbp^2+ \  \frac{ \om^2 \hbq^2}{2(1+\la \hbq^2)} =\frac{1}{2(1+\la \bq^2)}\big(-\hbar^2\De+\om^2\bq^2\big).
 \label{ca}
 \ee
Then:

\noindent
(i) $\hH$ commutes with the following observables:

\noindent
  $\bullet$ The $(2N-3)$ quantum angular momentum operators,
\be
  \hC^{(m)}=\!\! \sum_{1\leq i<j\leq m} \!\!\!\! ( \hq_i \hp_j- \hq_j \hp_i)^2  , \qquad 
  \hC_{(m)}=\!\!\! \sum_{N-m<i<j\leq N}\!\!\!\!\!\!  ( \hq_i \hp_j- \hq_j \hp_i)^2  ,  
  \label{cb}
 \ee
 where $m=2,\dots,N$ and $\hC^{(N)}=\hC_{(N)}$.
 
 \noindent
$\bullet$ The $N^2$ operators defining the ND quantum Fradkin tensor, given by
 \be
  \hI_{ij}=  \hp_i\hp_j- 2\la  \hq_i \hq_j{ \hH}( \hbq, \hbp)+ \om^2 \hq_i\hq_j , 
\label{cc}
\ee
where $i,j=1,\dots,N$ and   such that    ${\hH}=\frac 12 \sum_{i=1}^N \hI_{ii}$.

 \noindent
(ii) Each of the three  sets $\{{\hH},\hC^{(m)}\}$,  
$\{{\hH},\hC_{(m)}\}$ ($m=2,\dots,N$) and   $\{\hI_{ii}\}$ ($i=1,\dots,N$) is  formed by $N$ algebraically independent  commuting observables.

 \noindent
(iii) The set $\{ { \hH},\hC^{(m)}, \hC_{(m)},  \hI_{ii} \}$ for $m=2,\dots,N$ with a fixed index $i$    is  formed by $2N-1$ algebraically independent observables. 

 \noindent
(iv) $\hH$ is formally self-adjoint on the $L^2$ Hilbert space defined by the scalar product
\be
\langle \Psi | \Phi \rangle = \int_{\cM^N} \overline{{\Psi}(\bq)} \Phi(\bq)(1+\la\bq^2){\dd}\bq .
\label{product1}
\ee
\end{theorem}

 \noindent{\bf Proof.} Some points of this statement can be straightforwardly proven through the  coalgebra 
   symmetry~\cite{coalgebra1,coalgebra2,coalgebra3} of the quantum Hamiltonian (\ref{ca}). Let us consider the $\mathfrak{sl}(2,\RR)$ Lie coalgebra in the   basis $\{J_±,J_3\}$ with commutation rules, Casimir invariant  and (nondeformed) coproduct given by
 \be
 [J_3,J_+]=2\rmi \hbar J_+,\qquad  [J_3,J_-]=-2\rmi \hbar J_-,\qquad  [J_-,J_+]=4\rmi \hbar J_3,
 \label{vxa} 
 \ee
 \be
 {\cal C}=\frac 12 (J_+ J_- + J_- J_+)- J_3^2,
  \label{vxb} 
\ee
\be
\Delta(J_l)=J_l\otimes 1+ 1\otimes J_l,\qquad l=+,-,3.
 \label{vxc} 
\ee
 An $N$-particle realization of $\mathfrak{sl}(2,\RR)$ reads
 \be
 J_+=\hbp^2,\qquad J_-=\hbq^2,\qquad J_3=\frac 12(\hbq\cdot\hbp+\hbp\cdot\hbq)=\hbq\cdot\hbp -\frac 12 \rmi \hbar N .
  \label{vxd} 
 \ee
 Therefore, $\hH$ (\ref{ca}) has an  $\mathfrak{sl}(2,\RR)$ coalgebra symmetry since it can be written as
\be
 \hH= \frac{1}{2(1+\la J_-)}\, J_+ + \  \frac{ \om^2 J_-}{2(1+\la J_-)} .
\label{vxe}
\ee
 Hence, by construction,  $\hH$ commutes with the $(2N-3)$ observables $\hC^{(m)}$ and $\hC_{(m)}$ $(m=2,\dots,N)$ (\ref{cb}) which come  from the ``left" and ``right" $m$-th coproducts~\cite{coalgebra2,coalgebra3} of the invariant (\ref{vxb}), respectively, up to an additive constant  $\hbar^2m(m-4)/4$. Furthermore, the coalgebra approach also ensures that these are algebraically independent and that each set $\{{\hH},\hC^{(m)}\}$ and
$\{{\hH},\hC_{(m)}\}$ is formed by $N$ commuting observables (to be more precise, they are  polynomially independent as operators in a Jordan
algebra).

  Next, by direct computations it can be   proven that the $N^2$ observables $\hI_{ij}$ (\ref{cc}) commute with $\hH$, and that 
    the $N$ (diagonal)  observables  $\hI_{ii}$  $(i=1,\dots, N)$ commute amongst themselves as well;  it is obvious that the latter $\hI_{ii}$ are algebraically independent. Finally, it is also clear that any {single} $ \hI_{ii}$ is algebraically independent with respect  to the set  of $2N-2$ observables $\{ { \hH},\hC^{(m)}, \hC_{(m)}\}$ (as it is when $\la=0$) $\square$.
 \medskip
 
 We stress that, as a byproduct of the above proof, any quantum Hamiltonian defined as a function of (\ref{vxd}),
\be
\hH=\hH(J_+,J_-,J_3)=\hH(\hbp^2,\hbq^2,\hbq\cdot\hbp -  \rmi \hbar N/2), 
 \label{vxf}
 \ee
 is endowed with the same $\mathfrak{sl}(2,\RR)$ coalgebra symmetry.  This shows that this is quasi-MS~\cite{annals,coalgebra2,coalgebra3}, that is,  it commutes,  at least,  with the    
  $(2N-3)$ observables $\hC^{(m)}$ and $\hC_{(m)}$.    In this respect, we remark that what makes  the quantum Darboux III oscillator (\ref{vxe}) very special, is the existence of  a quantum Fradkin tensor formed by the ``additional" symmetries  $ \hI_{ij}$. This algebraic property implies that the system is MS and, as we shall see, that its discrete energy spectrum is maximally degenerate.     
 
 
 \subsection{The Laplace--Beltrami quantization}

When dealing with curved spaces with metric and classical kinetic term given by
$$
{\rm d}s^2=\sum_{i,j=1}^N g_{ij}(\bq){\rm d}q_i{\rm d}q_j,\qquad {\cal T}(\bq,\bp)=\frac12\sum_{i,j=1}^N g^{ij}(\bq) {p_ip_j}\, ,
$$
the  LB operator   
$$
\Delta_{\rm LB}=\sum_{i,j=1}^N \frac 1{\sqrt{g}}\partial_i\sqrt{g} g^{ij}\partial_j ,
$$
can be used in order  to define the quantum kinetic energy as 
$$
\hat{\cal T}_{\rm LB}(\hat\bq,\hat\bp)=-\frac {\hbar^2} 2 \Delta_{\rm LB},
\label{kg}
$$
where $g^{ij}$ is the inverse of the metric tensor $g_{ij}$ and $g$ is its determinant (see, for instance,~\cite{KKMW03,REFS1}).
If we apply such LB quantization to the Hamiltonian (\ref{ac}) with metric tensor~(\ref{metric1}) we get
\bea
&& \hat{\cal H}_{\rm LB}=-\frac {\hbar^2} 2 \Delta_{\rm LB}+ \frac{\omega^2 \mathbf{q}^2}{2(1+\lambda \mathbf{q}^2)} \nonumber\\
&&\qquad\, = -\frac{\hbar^2}{2(1+\lambda \mathbf{q}^2)}\De - \frac{\hbar^2\lambda(N-2)}{2 (1+\lambda\mathbf{q}^2)^2}\, (\mathbf{q}\cdot\nabla )+ \frac{\omega^2\mathbf{q}^2}{2(1+\lambda\mathbf{q}^2)} .\nonumber
\eea
Then,  $\hat {\cal H}$ (\ref{ca}) and $\hat{\cal H}_{\rm LB}$ only coincide in the case $N=2$ (as it should be   for any sperically symmetric space~\cite{Kalnins1}) and for $N>2$ they differ  by a momentum-dependent potential, namely:
$$
 \hat{\cal H}_{\rm LB}= \hat{\cal H}+\uu,\qquad \uu(\hat\bq, \hat\bp)= - {\rm i} \frac{\hbar\lambda(N-2)}{2 (1+\lambda\mathbf{\hat q}^2)^2} (\hat\bq\cdot\hat\bp) ,
$$
 where we have introduced the quantum variables (\ref{ka}). Notice that $\uu$ is linear in  $\hbar$, so this term does not have any classical analog. This situation reminds what happens in the context of the so-called {\emph{quasi-exactly solvable}} quantum models~\cite{REFS2}.  
 On the other hand, although the Hamiltonian  $ \hat{\cal H}_{\rm LB}$ commutes with the operators (\ref{cb}) (the quantum correction $\uu$  preserves the $\mathfrak{sl}(2,\RR)$ coalgebra symmetry (\ref{vxa})--(\ref{vxd})),  in this case there is no hint about the existence of an additional 
 symmetry of the type (\ref{cc}). 
 
 Nevertheless, it is possible to find a 
 ``superintegrable" LB quantization (in the sense that it does preserve the MS property) by adding a second potential term to $\hat {\cal H}$ (besides  $\uu$) thus conveying $N^2$  additional integrals of the type (\ref{cc}) together with the separability property in terms of the $N$ ``diagonal" ones. In order to achieve this result  we will relate $ \hat{\cal H}$ and $ \hat{\cal H}_{\rm LB}$ through a similarity transformation. If we apply 
 $ \hat{\cal H}_{\rm LB}$ to the product  $\exp(\ff(\bq))\Psi(\bq)$ we get:
\bea
\hat{\cal H}_{\rm LB}\rme^\ff\Psi \!\!&= &\!\! -\frac{\hbar^2 \rme^\ff}{2(1+\lambda \mathbf{q}^2)}\De \Psi -\frac{\hbar^2 \rme^\ff}{1+\lambda \mathbf{q}^2} (\nabla \ff \cdot \nabla \Psi) -\frac{\hbar^2 \rme^\ff}{2(1+\lambda \mathbf{q}^2)} (\De \ff + (\nabla \ff)^2)\Psi \nonumber\\
 &&  -\frac{\hbar^2 \lambda (N-2)\rme^\ff}{2(1+\lambda \mathbf{q}^2)^2}\,(\mathbf{q}\cdot\mathbf\nabla\Psi) -\frac{\hbar^2 \lambda (N-2)\rme^\ff}{2(1+\lambda \mathbf{q}^2)^2}(\mathbf{q}\cdot\mathbf\nabla \ff)\Psi + \frac{\omega \mathbf{q}^2}{2(1 + \lambda \mathbf{q}^2)}\rme^\ff\Psi .
 \nonumber
\eea
The two terms depending on $(\nabla \ff \cdot \nabla \Psi)$ and  $(\bq\cdot \nabla\Psi)$ can be removed by setting
\be
\ff(\bq) = \frac{2-N}{4}\ln (1+ \lambda \mathbf{q}^2)  ,
\label{ki}
\ee
which, in turn, means that there is a similarity transformation connecting $\hat {\cal H}$ and $\hat{\cal H}_{\rm LB}$:
\bea
&&
\hat{\cal H}_{\rm LB}\rme^\ff\Psi = \rme^\ff (\hat{\cal H}-\pot(\mathbf{q}))\Psi ,  \qquad  \hat{\cal H}_{\rm LB} = \rme^\ff \hat{\cal H}\rme^{-\ff }-\pot , \nonumber\\
&&     \pot(\mathbf{q})= -\frac{\hbar^2\la (N-2)}{8(1+\lambda \mathbf{q}^2)^3}\left( 2N+3\la\bq^2(N-2)\right)  .
\label{kj}
\eea
Notice that the multiplication operator $\rme^f$ defines a transformation mapping 
$$
L^2(\cM^N)=\left(\RR^N,(1+\la\bq^2)\,\dd \bq\right)\quad {\rm into}\quad L^2(\cM^N)= L^2\left(\RR^N,(1+\la\bq^2)^{N/2}\dd\bq\right),
$$
 which is the natural $L^2$ space defined by the Riemannian metric.
We remark that either for $N=2$ or when $\la=0$, $\ff(\bq)=\pot(\mathbf{q})=0$, and notice also that the central potential  $\pot$ is a pure quantum term as it depends on $\hbar^2$.
The latter result suggests to consider a ``transformed-LB" Hamiltonian defined by
\be
\hamG=\hat{\cal H}_{\rm LB}+\pot=\hat{\cal H} + \uu+\pot ,
\label{kk}
\ee
 which satisfies
 \be
 \hamG= \rme^\ff \hat {\cal H} \rme^{-\ff} .
 \label{kl}
  \ee
Hence, as a direct consequence, all the symmetries of $\hat {\cal H}$ give rise to those corresponding to    $\hamG$:
\be
\hat X_{\rm TLB}=\rme^\ff \hat X \rme^{-\ff} ,\qquad \hat X=    \{ \hC^{(m)},  \hC_{(m)},  \hI_{ij}\},\qquad [\hamG,\hat X_{\rm TLB}]=0 .
\label{km}
\ee
Therefore, by taking into account Theorem 2 and the equations (\ref{kl}) and (\ref{km}) we find that  $ \hamG$ is, in fact, a quantum MS Hamiltonian.

 \begin{theorem} Let   $\hamG$ be the quantum Hamiltonian given by
 \begin{align}
 \hamG &= 
 \frac{1}{2(1+\la \hbq^2)}\, \hbp^2+ \  \frac{ \om^2 \hbq^2}{2(1+\la \hbq^2)}  - {\rm i} \frac{\hbar\lambda(N-2)}{2 (1+\lambda\mathbf{\hat q}^2)^2} (\hat\bq\cdot\hat\bp) \nonumber\\
 &\qquad    -\frac{\hbar^2\la (N-2)}{8(1+\lambda  {\hbq}^2)^3}\left( 2N+3\la\hbq^2(N-2)\right)\nonumber \\
 &=-\frac{\hbar^2}2\De_{\rm LB}+\frac{\omega^2\mathbf{q}^2}{2(1+\lambda\mathbf{q}^2)} -\frac{\hbar^2\la (N-2)}{8(1+\lambda  {\bq}^2)^3}\left( 2N+3\la\bq^2(N-2)\right).
 \label{kn}
 \end{align}
Then:

\noindent
(i) $\hamG$ commutes with the  same   observables (\ref{cb}), that is,  $\hC_{\rm TLB}^{(m)}=\hC^{(m)}$  and  $\hC_{{\rm TLB},(m)}=\hC_{(m)}$,   as well as with the   $N^2$ Fradkin  operators  given by
  \bea
&&  \hI_{{\rm TLB},ij}= \hp_i\hp_j -(N-2)\frac{\rmi\hbar\la}{2(1+\la\hbq^2)} \,(\hq_i\hp_j +\hq_j\hp_i ) + \frac{(N-2)\hbar^2\la^2\hq_i\hq_j}{(1+\la\hbq^2)^2}\left(1-\frac{N-2}{4} \right) \nonumber   \\ 
 &&\qquad \quad \qquad  - \frac{(N-2)\hbar^2\la}{2(1+\la\hbq^2)}\,\delta_{ij}
   - 2\la  \hq_i\hq_j { \hamG}( \hbq, \hbp)+ \om^2 \hq_i\hq_j ,
\label{kp}
\eea
with $ i,j=1,\dots,N$ and such that  ${\hamG}=\frac 12 \sum_{i=1}^N  \hI_{{\rm TLB},ii}$.  

\noindent
(ii) Each of the three  sets $\{{\hamG},\hC^{(m)}\}$,  
$\{{\hamG},\hC_{(m)}\}$ ($m=2,\dots,N$) and   $\{  \hI_{{\rm TLB},ii}\}$ ($i=1,\dots,N$) is  formed by $N$ algebraically independent  commuting observables.

\noindent
(iii) The set $\{ { \hamG},\hC^{(m)}, \hC_{(m)},    \hI_{{\rm TLB},ii}\}$ for $m=2,\dots,N$ with a fixed index $i$    is  formed by $2N-1$ algebraically independent observables.

\noindent
(iv) $\hamG$ is formally self-adjoint on the space $L^2(\cM^N)$ associated with the underlying Darboux space, defined by
\be
\langle \Psi | \Phi \rangle_{\rm TLB} = \int_{\cM^N} \overline{{\Psi}(\bq)}\, \Phi(\bq)\, (1+\la\bq^2)^{N/2}\,\dd\bq .
\label{product2}
\ee
\end{theorem}
 
 Therefore, according to the above statement, $ \hamG$ can be seen as the  appropriate  LB-quantization of the classical Hamiltonian (\ref{ac}) as it manifestly preserves the MS property. Such a quantization    requires to add a linear momentum-dependent potential $\uu$  (coming from the quantum kinetic energy) plus an ``additional" central potential  $\pot$ (coming from the MS property) to the Hamiltonian $\hat {\cal H}$. Clearly, the eigenfunctions of  $\hamG$ can be read off from  those of $\hat {\cal H}$  by means of (\ref{kl}).

\noindent {\bf Remark.} 
Interestingly, the quantum correction  $\pot$ (\ref{kj}) to the oscillator potential arising from the similarity transformation 
 (\ref{ki}) is proportional to the scalar curvature of the underlying metric (\ref{ad}):
 $$
\pot=\frac{\hbar^2 (N-2)}{8(N-1)} \,R  ,
$$
which vanishes for any  2D space and in the (most frequently studied) case of spaces of constant curvature gives simply an additional constant.
Therefore, the transformed-LB quantization prescription is equivalent to imposing that
$$
\hamG=-\frac{\hbar^2}2 \Delta_{\rm c}+\mathcal U(\bq)\,,
$$
that is, to asserting that the appropriate quantum kinetic energy operator is essentially the {conformal Laplacian} (see, for instance,~\cite{Baer})
$$
  \Delta_{\rm c} = \Delta_{\rm LB} -  \frac{  (N-2)}{4(N-1)} \,R     \, ,
  $$
  rather than the ordinary Laplacian (or LB operator).
  This is in full agreement with many prescriptions used in the analysis of scalar field theories in General Relativity or when dealing with quantization on arbitrary Riemannian manifolds~\cite{Wa84, Landsman, Liu}. However, we are not aware of any other instances where the convenience of this prescription has been motivated by superintegrability arguments.

 
 \subsection{A position-dependent mass quantization}

In the framework  of PDM Hamiltonian systems there are also several ways to define the quantum kinetic energy term. A general approach depending on three parameters subjected to a constraint can be found in~\cite{Roos} (see also~\cite{Quesnea}). We shall consider here the proposal given in~\cite{mass2} and based on Galilean invariance arguments, which is the one extensively used in the condensed matter literature~\cite{Bastard, qDWW}. Such a PDM quantization, ${\cal T}(\bq,\bp)\to \hat{\cal T}_{\rm PDM}(\hat\bq,\hat\bp)$, is defined as
$$
  \hat{\cal T}_{\rm PDM}(\hat\bq,\hat\bp)=\frac { 1} 2\, \hbp\cdot \frac{1}{(1+\la \hbq^2)} \hbp=-\frac {\hbar^2} 2 \nabla\cdot  \frac{1}{(1+\la \bq^2)} \nabla  .
\label{oa}
$$
Then, by adding the oscillator potential and ordering terms in the kinetic term, we obtain the following PDM quantization of the 
 Hamiltonian (\ref{ac}):
\bea
 && \hat{\cal H}_{\rm PDM}=  \hat{\cal T}_{\rm PDM}(\hat\bq,\hat\bp)+{\cal U}(\hbq)\nonumber\\
  &&\qquad \quad =
     -\frac{\hbar^2}{2(1+\lambda \mathbf{q}^2)}\De + \frac{\hbar^2\lambda }{(1+\lambda\mathbf{q}^2)^2}\, (\mathbf{q}\cdot\nabla )+ \frac{\omega^2\mathbf{q}^2}{2(1+\lambda\mathbf{q}^2)} .
\nonumber
 \eea
Hence  the difference between $\hat {\cal H}$ (\ref{ca}) and $\hat{\cal H}_{\rm PDM}$  relies again in a momentum-dependent potential,  
$$
 \hat{\cal H}_{\rm PDM}= \hat{\cal H}+\vv,\qquad \vv(\hat\bq, \hat\bp)=  {\rm i} \frac{\hbar\lambda }{ (1+\lambda\mathbf{\hat q}^2)^2} (\hat\bq\cdot\hat\bp) .
$$
Similarly to the LB quantization,  the MS property can be explicitly restored through a similarity transformation and this process will require to add another central potential to the initial $ \hat{\cal H}_{\rm PDM}$.
 
 Explicitly, if we apply  
 $ \hat{\cal H}_{\rm PDM}$ to the product  $\exp(\hh(\bq))\Psi(\bq)$ and define
$$
\hh(\bq) = \frac{1}{2}\ln (1+ \lambda \mathbf{q}^2)  ,
$$
then we get the following similarity transformation   between $\hat {\cal H}$ and $\hat{\cal H}_{\rm PDM}$:
\bea
&&
\hat{\cal H}_{\rm PDM}\rme^\hh\Psi = \rme^\hh (\hat{\cal H}-\vpot(\mathbf{q}))\Psi ,  \qquad  \hat{\cal H}_{\rm PDM} = \rme^\hh \hat{\cal H}\rme^{-\hh }-\vpot , \nonumber\\
&&     \vpot(\mathbf{q})= \frac{\hbar^2\la  }{2(1+\lambda \mathbf{q}^2)^3}\left( N+\la\bq^2(N-3)\right)  .
\nonumber
\eea
Hence, in contrast with the LB quantization, now  both $\hh(\bq)$ and $\vpot(\bq)$ are nontrivial for any dimension $N$ (including $N=2$). In this way, we define the following ``transformed-PDM" Hamiltonian,
\be
\hat{\cal H}_{\rm TPDM}=\hat{\cal H}_{\rm PDM}+\vpot=\hat{\cal H} + \vv+\vpot ,\qquad  \hat{\cal H}_{\rm TPDM}= \rme^\hh \hat {\cal H} \rme^{-\hh},
\label{oe}
\ee
whose symmetries are thus obtained from those of $ \hat {\cal H} $ as
\be
\hat X_{\rm TPDM}=\rme^\hh \hat X \rme^{-\hh} ,\qquad \hat X=    \{ \hC^{(m)},  \hC_{(m)},  \hI_{ij}\},\qquad [\hat{\cal H}_{\rm TPDM},\hat X_{\rm TPDM}]=0 .
\label{of}
\ee

The MS property of the Hamiltonian $\hat{H}_{\rm TPDM}$ is summarized in the following statement.
  
 \begin{theorem} Let   $\hat{\cal H}_{\rm TPDM}$ be the quantum Hamiltonian defined by
 \begin{align}
 \hat{\cal H}_{\rm TPDM} = 
 \frac{1}{2(1+\la \hbq^2)}\, \hbp^2+ \  \frac{ \om^2 \hbq^2}{2(1+\la \hbq^2)}  +  \frac{{\rm i}\hbar\lambda }{ (1+\lambda\mathbf{\hat q}^2)^2} (\hat\bq\cdot\hat\bp)+ \frac{\hbar^2\la \left( N+\la\bq^2(N-3)\right) 
  }{2(1+\lambda \mathbf{q}^2)^3} .
 \label{og}
 \end{align}
 Then:
 
\noindent
(i) $\hat{\cal H}_{\rm TPDM} $ commutes with the      observables (\ref{cb}) as well as with    $( i,j=1,\dots,N)$
\bea
&&  \hI_{{\rm TPDM},ij}= \hp_i\hp_j +\frac{\rmi\hbar\la}{(1+\la\hbq^2)} \,(\hq_i\hp_j +\hq_j\hp_i) + \frac{\hbar^2\la }{(1+\la\hbq^2)}\left(\delta_{ij}-\frac{3 \la \hq_i\hq_j }{(1+\la \hbq^2)} \right)\nonumber\\
&&\qquad \qquad \quad    - 2\la  \hq_i\hq_j { \hat{\cal H}_{\rm TPDM} }( \hbq, \hbp)+ \om^2 \hq_i\hq_j,
   \nonumber
\eea
 which form a quantum Fradkin tensor and verifiy that $\hat{\cal H}_{\rm TPDM} =\frac 12 \sum_{i=1}^N  \hI_{{\rm TPDM},ii}$.  

\noindent
(ii) Each of the three  sets $\{{ \hat{\cal H}_{\rm TPDM} },\hC^{(m)}\}$,  
$\{{ \hat{\cal H}_{\rm TPDM} },\hC_{(m)}\}$ ($m=2,\dots,N$) and   $\{  \hI_{{\rm TPDM},ii}\}$ ($i=1,\dots,N$) is  formed by $N$ algebraically independent  commuting observables.

\noindent
(iii) The set $\{ {  \hat{\cal H}_{\rm TPDM} },\hC^{(m)}, \hC_{(m)},    \hI_{{\rm TPDM},ii}\}$ for $m=2,\dots,N$ with a fixed index $i$    is  formed by $2N-1$ algebraically independent observables.

\noindent
(iv) $ \hat{\cal H}_{\rm TPDM} $ is formally self-adjoint on the standard $L^2$ space with product
\[
\langle \Psi | \Phi \rangle_{\rm TPDM }= \int_{\cM^N} \overline{{\Psi}(\bq)}\, \Phi(\bq)\,  \dd\bq .
\]
\end{theorem}

\bigskip

Finally, we remark that by combining the similarity transformations  (\ref{kl}) and (\ref{oe}) we obtain the relationship between $\hamG$ and $\hat{\cal H}_{\rm TPDM}$:
$$
\hat{\cal H}_{\rm TPDM}=\rme^{\hh-\ff} \hamG \rme^{-(\hh-\ff)} =(1+\la \bq^2)^{N/4} \hamG(1+\la \bq^2)^{-N/4} .
$$

 
 \sect{Radial Schrödinger equations}

In this section we obtain the 1D radial Schrödinger equation coming from each of the above three $N$D quantum 
Hamiltonians by,  firstly, introducing  
hyperspherical coordinates and, secondly, by making use of the observables $\hC_{(m)}$ ({\ref{cb}) that encode the full spherical symmetry of the three systems.

Let us   introduce the map  from the initial quantum operators   (\ref{ka}) to the quantum  hyperspherical ones $\hr$, $\hte_j$, $\hp_r$, $\hp_{\te_j}$ ($j=1.\dots,N-1)$ with   Lie brackets and differential representation given by
 \bea
&&\!\!\!\!\!\!\! [ \hr,\hp_r]=\rmi \hbar, \qquad [ \hr,\hp_{\te_j}]=0, \qquad [\hte_j,\hp_r]=0, \qquad  [\hte_j,\hp_{\te_k}]=\rmi \hbar\delta_{jk},\nonumber\\
&&\!\!\!\!\!\!\! \hr=r,\qquad  \hp_r= -\rmi\hbar\frac{\partial}{\partial r} ,\qquad \hte_j=\te_j,\qquad \hp_{\te_k}=-\rmi\hbar \frac{\partial}{\partial{\te_j}}.
\label{la}
\eea
Here we point out that the ``radial and phase operators"  that we have just introduced are nothing but formal multiplicative operators on the angular variables, 
whose ``canonical" transformation rules with respect to the Cartesian ones are: 
 \bea
&&\hq_j=\hr \cos\hte_{j}     \prod_{k=1}^{j-1}\sin\hte_k, \quad 1\leq j<N;\qquad 
\hq_N =\hr \prod_{k=1}^{N-1}\sin\hte_k ,\nonumber\\
&& \hp_j=\prod_{k=1}^{j-1}\sin\hte_k\cos\hte_j\,\hp_r+\frac{\cos\hte_j}{\hr}\sum_{l=1}^{j-1} \frac{\prod_{k=l+1}^{j-1}\sin\hte_k }{\prod_{m=1}^{l-1}\sin\hte_m} \cos\hte_l\,\hp_{\te_l}-\frac{\sin\hte_j}{\hr\prod_{k=1}^{j-1}\sin\hte_k}\,\hp_{\te_j} , \nonumber\\
&&\hp_N=\prod_{k=1}^{N-1}\sin\hte_k \,\hp_r+\frac 1 \hr\sum_{l=1}^{N-1}\frac{\prod_{k=l+1}^{N-1}\sin\hte_k}{\prod_{m=1}^{l-1}\sin\hte_m}\cos\hte_l \, \hp_{\te_l} .
\nonumber
\eea
Hence we  obtain that
  \begin{equation}
  \hbq^2=\hr^2 , \quad
\hbp^2=\frac{1}{\hr^{N-1}}\, \hp_r\, \hr^{N-1} \,\hp_r+\frac{\hbL^2}{\hr^{2}} =\hp_r^2 - \rmi \hbar\frac{(N-1)}{\hr}\, \hp_r  +\frac{\hbL^2}{\hr^{2}} ,\quad \hbq\cdot\hbp=\hr \hp_r,
\label{lc}
\end{equation}
where $\hbL^2$ is the square of the total  quantum  angular momentum    given by
$$
\hbL^2=\sum_{j=1}^{N-1}\left( \prod_{k=1}^{j-1}\frac{1}{\sin^{2}\hte_k}  \right) \frac{1}{(\sin\hte_j)^{N-1-j}}\, \hp_{\te_j}(\sin\hte_j)^{N-1-j} \,\hp_{\te_j}.
\label{ld}
$$
Notice that the expressions (\ref{lc}) provide a 1D (radial) representation of the $\mathfrak{sl}(2,\RR)$ Lie algebra (\ref{vxa}) by introducing them in 
(\ref{vxd}).
 
The $N-1$ commuting observables $\hC_{(m)}$ ({\ref{cb}) turn out to be $(m=2,\dots,N)$
$$
\hC_{(m)}=\sum_{j=N-m+1}^{N-1}\left(\prod_{k=N-m+1}^{j-1}\frac 1{\sin^{2}\hte_k  }  \right) \frac{1}{(\sin\hte_j)^{N-1-j}}\, \hp_{\te_j} (\sin\hte_j)^{N-1-j} \,  \hp_{\te_j}  ,
\label{le}
$$
with $\hC_{(N)} =\hbL^2$.  Thus  we obtain a  set of $N-1$ angular  equations  $(k=3,\dots,N-1)$:
\bea
&&\!\!\!\! \hC_{(2)}(\hte_{N-1},\hp_{\te_{N-1}})=\hp^2_{\te_{N-1}} , \nonumber\\[2pt]
&&\!\!\!\!  \hC_{(k)}(\hte_{N-k+1},\hp_{\te_{N-k+1}})=\frac{1}{(\sin\hte_{N-k+1})^{k-2}} \, \hp_{\te_{N-k+1}} (\sin\hte_{N-k+1})^{k-2} \hp_{\te_{N-k+1}}  +
\frac{\hC_{(k-1)} }{\sin^{2}\hte_{N-k+1}} ,  \nonumber \\[2pt]
&&\!\!\!\!  \hC_{(N)}(\hte_{1},\hp_{\te_{1}})=\frac{1}{(\sin\hte_{1})^{N-2}} \, \hp_{\te_{1}} (\sin\hte_{1})^{N-2} \,\hp_{\te_{1}} +
\frac{\hC_{(N-1)} }{\sin^{2}\hte_1}\equiv  \hbL^2 ,
\label{lf}
\eea
which are worth to be compared with (\ref{bi}). Therefore the quantum radial Hamiltonian corresponding to (\ref{ca}) is obtained in the form
  \be
 {\hH}(\hr , \hp_r\ )=  \frac{1 }{2(1+\la \hr^2)}\left( \frac{1}{\hr^{N-1}}\, \hp_r\, \hr^{N-1} \,\hp_r+\frac{\hbL^2}{\hr^{2}}+\om^2 \hr^2 \right)    .
 \label{lg}
 \ee
 After reordering   terms and  introducing  the differential operators (\ref{la})  in the Hamiltonian (\ref{lg}) we arrive at the following Schrödinger equation, $ {\hH}\Psi=E\Psi$,  
      \be
\frac{1 }{2(1+\la r^2)}\left(  {-\hbar^2}\partial_r^2 -\frac{\hbar^2(N-1)}{r}\,\partial_r + \frac{\hat{\mathbf{L}}^2}{r^2} + \omega^2 r^2  \right)\Psi(r, \boldsymbol{\te}) = E\Psi(r,{\boldsymbol{\te}}) ,
 \label{lh}
 \ee
where $\boldsymbol{\te}=(\te_1,\dots,\te_{N-1})$.  Next we factorize   the  wave function in the usual  radial and angular components  and consider 
    the separability provided by the  first integrals   $\hat{C}_{(m)}$ (\ref{lf})  with eigenvalue equations given by
\be
\Psi(r, \boldsymbol{\te})= \phiS(r)Y(\boldsymbol{\te}),\qquad \hat{C}_{(m)}\Psi = c_{m} \Psi,\quad m=2,\dots , N.
\label{rf}
\ee
Consequently, 
we obtain that $Y(\boldsymbol{\te})$ solves completely  the angular part and such hyperspherical harmonics verify
\be
 \hat{C}_{(N)} Y(\boldsymbol{\te})=   \hbL^2 Y(\boldsymbol{\te})=\hbar^2 l(l+N-2)Y(\boldsymbol{\te}),\quad l=0,1,2\dots  
\label{rg}
\ee
where $l$ is the angular quantum number. By taking into account  the angular equations (\ref{lf}), we find that the eigenvalues  $c_m$ of the operators  $\hat{C}_{(m)}$ are related to the  $N-1$ quantum numbers of the angular observables as 
$$
c_{k}\leftrightarrow l_{k-1},\quad k=2,\dots, N-1,\qquad c_N \leftrightarrow l ,
$$
 that is, 
$$
Y(\boldsymbol{\te})\equiv Y^{c_N}_{c_{N-1},..,c_2}(\theta_1,\theta_2,...,\theta_{N-1}) \equiv Y^{l}_{l_{N-2},..,l_1}(\theta_1,\theta_2,...,\theta_{N-1}) .
$$
 Hence the radial Schrödinger equation provided by $\hat{\cal H}$  is
\be
\frac 1{2(1+\la r^2)} \left( {-\hbar^2}\left(\frac{\dd^2}{\dd r^2} +\frac{(N-1)}{r}\frac{\dd}{\dd r} -\frac{l(l+N-2)}{r^2}\right)+\om^2 r^2\right) \phiS(r) = E \phiS(r) .
\label{sa}
\ee

In the same way, the 1D radial Hamiltonian operators coming from the transformed LB (\ref{kn}) and PDM  (\ref{og}) quantizations are found to be
\bea
 \hamG \!\!&= &\! \! - \frac {\hbar^2}{2(1+\la r^2)} \left( \frac{\dd^2}{\dd r^2} +\left( \frac{N-1}{r}   +\frac{\la (N-2)r}{1+\la r^2} \right) \frac{\dd}{\dd r} -\frac{l(l+N-2)}{r^2}\right)\nonumber\\
 &&\quad +\frac{\om^2 r^2}{2(1+\la r^2)} -\frac{\hbar^2\la (N-2)}{8(1+\lambda  {r}^2)^3}\left( 2N+3\la r^2(N-2)\right),
 \label{li}\\[4pt]
\hat{\cal H}_{\rm TPDM} \!\!&= &\! \! - \frac {\hbar^2}{2(1+\la r^2)} \left( \frac{\dd^2}{\dd r^2} +\left( \frac{N-1}{r}   -\frac{2\la r}{1+\la r^2} \right) \frac{\dd}{\dd r} -\frac{l(l+N-2)}{r^2}\right)\nonumber\\
 &&\quad +\frac{\om^2 r^2}{2(1+\la r^2)}+ \frac{\hbar^2\la \left( N+\la r^2(N-3)\right) 
  }{2(1+\lambda r^2)^3}.
 \label{lj}
 \eea

Recall that the three radial Hamiltonians $ {\hat {\cal H}}$, $\hamG$ and $ \hat{\cal H}_{\rm TPDM}$,  are related through the similarity transformations as
\bea
&&\hamG=(1+\la r^2)^{(2-N)/4} {\hat {\cal H}} (1+\la r^2)^{(N-2)/4},\nonumber\\
&&  \hat{\cal H}_{\rm TPDM}= (1+\la r^2)^{1/2} \hat {\cal H}(1+\la r^2)^{-1/2},\nonumber\\
&& \hat{\cal H}_{\rm TPDM}= (1+\la r^2)^{N/4} \hamG(1+\la r^2)^{-N/4} 
.\nonumber
\eea
Therefore the three corresponding radial Schrödinger equations share the same energy spectrum and have different but equivalent radial wave functions:
\bea
&&  \hat{\cal H} \phiS(r) =E\Phi(r) ,\quad   \hamG \phiLB(r) =E\phiLB(r) ,\quad 
\hat{\cal H}_{\rm TPDM}\phiPDM(r) =E\phiPDM(r),\nonumber\\
&& \phiLB(r)=(1+\la r^2)^{(2-N)/4} \phiS(r),\quad   \phiPDM(r)=(1+\la r^2)^{1/2} \phiS(r),\nonumber\\
&& \phiPDM(r)=(1+\la r^2)^{N/4} \phiLB(r).\label{lk}
\eea

 
 \sect{Spectrum and eigenfunctions}

In this section we shall compute, in a rigorous manner,   the (continuous and discrete) spectrum and eigenfunctions of the quantum nonlinear oscillator by using the quantum Hamiltonian $\hamG$ (\ref{og}) characterized in Theorem 3. We recall that the results corresponding to the Schrödinger quantization $\hat{\cal H}$ (\ref{ca}) of Theorem 2 were   advanced in~\cite{FTC} but without   explicit proofs. Recall that both quantizations share the same spectrum but they have different  
radial wave functions which are related through the similarity transformation (\ref{lk}).

 
 \subsect{Continuous spectrum}

Since $\cM^N$ is a complete manifold and the potential is continuous and bounded, it is standard that $\hamG$ is essentially self-adjoint on the space $C^\infty_0(\RR^N)$ of smooth functions of compact support. It should be remarked that one cannot immediately determine the continuous spectrum of $\hamG$ from asymptotics of the potential: in a complete Riemannian manifold, even the spectrum of the LB operator can be extremely difficult to analyze; 
e.g., it can be either purely continuous (as in Euclidean space), purely discrete~\cite{DL79} or consist of both a continuous part and eigenvalues, possibly embedded in the continuous spectrum~\cite{Do81}.

In fact, to compute the continuous spectrum of $\hamG$ it is convenient to take advantage of the spherical symmetry to decompose
\begin{equation}\label{l}
L^2(\cM^N)=\bigoplus_{l\in\NN}L^2(\RR^+,\dd\nu)\otimes\cY_l\,,
\end{equation}
where $\dd\nu(r)=r^{N-1}(1+\la r^2)^{N/2}\dd r$ and $\cY_l$ is the finite-dimensional space of (generalized) spherical harmonics, defined by
\[
\cY_l:=\big\{Y\in L^2(\SS^{N-1}):\De_{\SS^{N-1}}Y=-l(l+N-2)Y\big\}\, ,
\]
where $\NN$ stands for the set of nonnegative integers and  $\De_{\SS^{N-1}}$ denotes the Laplacian on the $(N-1)$D sphere $\SS^{N-1}$ (or minus the angular momentum operator).
This decomposition is tantamount to setting
\[
\Psi_\LB(\bq)=\sum_{l\in\NN}Y_l(\boldsymbol{\theta})\,\Phi_{\LB,l}(r)\,,
\]
with $\boldsymbol{\theta}=\bq/r\in\SS^{N-1}$, $r=|\bq|$ and $Y_l\in\cY_l$. 

As $\hamG$ is spherically symmetric, the decomposition~\eqref{l} allows us to write $\hamG$ as the direct sum of operators  
\begin{equation}\label{Hl}
\hamG=\bigoplus_{l\in\NN} \hat H_{\LB,l}\otimes \id_{\cY_l}\,,
\end{equation}
with each $\hat H_{\LB,l}$ standing for the Friedrichs extension of the differential operator on $L^2(\RR^+,\dd\nu)$; namely
\begin{multline*}
  2\hat H_{\LB,l}=-\frac{\hbar^2}{r^{N-1}(1+\la r^2)}\frac{\dd}{\dd r}r^{N-1}\frac{\dd}{\dd r}-\frac{\hbar^2\la (N-2)r}{(1+\la r^2)^2}\frac{\dd}{\dd r} +\frac{\hbar^2l(l+N-2)}{r^2(1+\la r^2)}\\
  +\frac{\om^2r^2}{1+\la r^2}-\frac{\hbar^2\la (N-2)}{4(1+\lambda  r^2)^3}\left( 2N+3\la r^2(N-2)\right) .
\end{multline*}
The continuous spectrum of $\hamG$ is most easily dealt with using this decomposition. Indeed, from~\eqref{Hl} it is apparent that
\[
\spec(\hamG)=\overline{\bigcup_{l\in\NN}\spec(\hat H_{\LB ,l})}\,.
\]

To understand the spectrum of $\hat H_{\LB ,l}$ we proceed to compute and analyze its associated quantum effective potential $\hat {\cal U}_{{\rm eff},l}$. For this purpose we      apply the same change of variable $Q=Q(r)$ (\ref{bbhh}) used in the classical case, together with  a change   of     the radial wave function $\Phi_{\LB,l}(r) \mapsto \ur(Q(r)) $. We require that these transformations    map the Schrödinger equation $\hat H_{\LB ,l}\Phi_{\LB,l}=E \Phi_{\LB,l}$ into
\bea
 \left( - \frac {\hbar^2}2 \frac{\dd^2}{\dd Q^2} + \hat {\cal U}_{{\rm eff},l}(  Q )\right) \ur(Q)=E \ur(Q).
 \label{sde}
\eea
 This is achieved by setting
\bea
\Phi_{\LB,l}(r) =\frac{r^{(1-N)/2}}{(1+\la r^2)^{(N-1)/4}}\, \ur(r)
\label{sc}
\eea
in the  radial Schrödinger equation,  thus yielding 
 \be
 \hat {\cal U}_{{\rm eff},l}(r  )=\frac{1}{2(1+\la r^2)}\left( \frac{\hbar^2\left(8(1+\la r^2)-5 \right)}{4r^2 (1+\la r^2)^2}+\frac{\hbar^2}{r^2}\left( l(l+N-2) + \frac{N(N-4)}4\right)+\om^2 r^2 \right) .
 \label{se}
\ee

The behavior of $ \hat {\cal U}_{{\rm eff},l}$ is rather similar to that of the classical effective potential (\ref{bbgg}) (see figure~\ref{figure4}), that is,
 $ \hat {\cal U}_{{\rm eff},l}$
 is   a positive function with a unique minimum, whose expression is rather cumbersome and which for the harmonic oscillator reduces to
 \bea
  \la=0:&&  r^2_{\rm min}=\hbar \frac{ \sqrt{  l(l+N-2) +(N-1)(N-3)/4}}{\om},\nonumber\\
  &&    \hat{\cal U}_{{\rm eff},l}(r_{\min})=\hbar \om \sqrt{    l(l+N-2) +(N-1)(N-3)/4} .
 \label{sf}
 \eea
 Similarly to the classical system,    the values of   $r_{\rm min}$ and $ \hat {\cal U}_{{\rm eff},l}(r_{\min})$   are  respectively  greater and smaller  than  those corresponding to the quantum harmonic oscillator  (\ref{sf}), but $ \hat {\cal U}_{{\rm eff},l}$ has the same asymptotic behaviour,  namely,
 \be
\lim_{r\to 0} \hat{\cal U}_{{\rm eff},l}(r)=+\infty,\qquad \lim_{r\to \infty} \hat{\cal U}_{{\rm eff},l}(r)=\frac{\om^2}{2\la}    .
\label{bzz}
\ee

We remark that there is a single exceptional  particular case for $l=0$ and $N=2$ for which $ \hat {\cal U}_{{\rm eff},l}$ reads
$$
 \hat {\cal U}_{{\rm eff},l}(r)=\frac{1}{2(1+\la r^2)}\left( \frac{-\hbar^2\left(1+4\la^2 r^4 \right)}{4r^2 (1+\la r^2)^2}+\om^2 r^2 \right) .
 \label{sea}
$$
Thus  $\lim_{r\to 0} \hat {\cal U}_{{\rm eff},l}=-\infty$ and  $\lim_{r\to \infty} \hat {\cal U}_{{\rm eff},l}=\omega^2/(2\la)$, so $ \hat {\cal U}_{{\rm eff},l}$ has no minimum and  can take both negative and positive values.


\begin{figure}
\includegraphics[height=7cm,width=10cm]{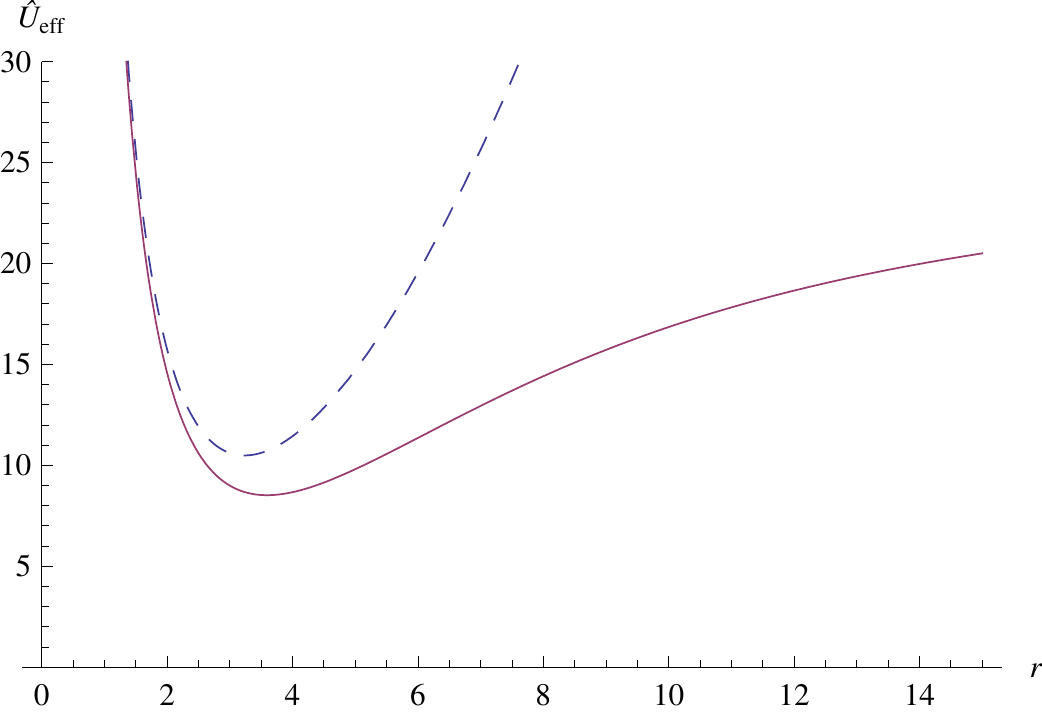}
\caption{The  {quantum effective  nonlinear   oscillator potential}  (\ref{se}) for  $N=3$,  $\la=0.02$, $l=10$ and $\hbar=\om=1$. The minimum of the potential is located at $r_{\rm min}=3.59$ with $\hat{\cal U}_{{\rm eff},l}(r_{\rm min})=8.52$ and $\hat{\cal U}_{{\rm eff},l}(\infty)=25$.  The dashed  line   corresponds to the quantum effective potential of the    isotropic oscillator  with $\la=0$ with minimum $\hat {\cal U}_{{\rm eff},l}(r_{\rm min})=10.49$ at  $r_{\rm min}=3.24$. 
 \label{figure4}}
\end{figure}


Since we have just related the nonnegative, self-adjoint second-order differential operator on the half-line $\hat H_{\LB ,l}$ to~\eqref{sde}, standard results in spectral theory~\cite[Theorem XIII.7.66]{DS88} ensure that the eigenvalues of $\hat H_{\LB ,l}$ are contained in $(0,E_\infty)$ and its continuous spectrum is absolutely continuous and given by
$[E_\infty,\infty)$, where we have set
\begin{align*}
  E_\infty&=\lim_{r\to\infty}\hat {\cal U}_{{\rm eff},l}=\frac{\om^2}{2\la}\,.
\end{align*}
Altogether, this guarantees that the continuous spectrum of $\hamG$ is
$$
\spec_{\rm cont}(\hamG)=\left[ {\om^2}/({2\la }),\infty\right),
\label{cont}
$$
 and that there are no embedded eigenvalues.

 
 \subsect{Discrete spectrum and eigenfunctions}

Let us now compute the eigenvalues and eigenfunctions of $\hamG$. To begin with, let us denote by $\psi_n(q)$ the $n$th eigenfunction of the 1D harmonic oscillator which satisfies
\[
\frac 12 \left(-\hbar^2\frac{\dd^2}{\dd q^2}+\om^2q^2 \right)\psi_n(q)= \hbar\om\left(n+\frac 12\right)\psi_n(q) \,.
\]
The explicit expression of $\psi_n$ in terms of Hermite polynomials is
\begin{equation}\label{Psi}
\psi_n(q)=\exp\left( - \frac {\om}{2\hbar} \,  q^2 \right)   \,H_n\left(\sqrt{\frac{\om}{\hbar} } \,q\right) ,
\end{equation}
up to a normalization constant.

Due to the relationship between the Schrödinger and LB quantizations~\eqref{kl} we have that $\Psi_\LB(\bq)=(1+\la \bq^2)^{(2-N)/4}\Psi(\bq)$ and the eigenvalue equation 
$$
\hamG\Psi_\LB(\bq)= E\Psi_\LB(\bq)
$$
can also be written as (see~(\ref{ca}))
\begin{equation}\label{phi}
  (-\hbar^2\De+\Om^2\bq^2)\Psi(\bq)=2E\Psi(\bq)\,,
\end{equation}
  where
\be
\Om=\sqrt{\om^2-2 \la E}.
\label{freq}
\ee
Since $\hamG$ has no embedded eigenvalues (as shown in   the previous subsection), one can safely assume that $\om^2-2 \la E>0$. The condition $\Psi_\LB\in L^2(\cM^N)$ translates, according to (\ref{product1}),  as
\[
\int |\Psi(\bq)|^2(1+\la \bq^2)\,\dd\bq<\infty\, ;
\]
in particular, $\Psi$ is square-integrable with respect to the Lebesgue measure.
Therefore, by the standard theory of the harmonic oscillator, there must exist some $n\in \NN$ such that
\[
E=\hbar\Om\bigg(n+\frac N2\bigg)\,.
\]
Substituting the formula for $\Om$, taking squares and isolating $E$, one readily finds that any eigenvalue of $\hamG$ must be of the form
\bea
E_n \!\!\!&=&\!\!\! - \lambda \hbar^2\left(n + \frac{N}{2}\right)^2 + \hbar \left(n+\frac N 2\right ) \sqrt{\hbar^2 \lambda^2 \left(n+\frac{N}{2}\right)^2+ \om^2 } 
\nonumber\\
\!\!\!&=&\!\!\!  \la\hbar^2 \bigg(n+\frac N2\bigg)^2\left(\sqrt{1+\frac{\om^2}{\hbar^2\la^2(n+\frac N2)^2}}-1\right)\,. 
\label{lan}
\eea
Conversely, one can prove that $E_n$ is an eigenvalue of $\hamG$ for any $n\in \NN$. This is easily seen by taking any partition $(n_i)_{i=1}^N\subset\NN$ such that $n_1+\cdots +n_N=n$ and noticing that, by~\eqref{Psi} and~\eqref{phi},
\begin{equation}\label{psin}
\Psi_\LB(\bq)=(1+\la \bq^2)^{(2-N)/4}\prod_{i=1}^N   \exp\{-\bb^2 q_i^2/2\} H_{n_i}(\bb q_i) ,\quad \bb=\sqrt{\frac{\omm}{\hbar}},
\end{equation}
is an $L^2(\cM^N)$ solution of the equation $\hamG\Psi_\LB=E_n\Psi_\LB$.

Together with the result of the previous subsection, this proves the following

\begin{theorem}\label{T.eigen} Let $\hamG$ be the quantum Hamiltonian   (\ref{og}). Then: 

\noindent
(i) The continuous spectrum of $\hamG$ is given by $[\frac{\om^2}{2\la },\infty)$. Moreover, there are no embedded eigenvalues and its singular spectrum is empty. 

\noindent
(ii)   $\hamG$ has an infinite number of eigenvalues, all of which are contained in $(0,\frac{\om^2}{2\la })$. Their only accumulation point is $\frac{\om^2}{2\la }$, that is, the bottom of the continuous spectrum. 

\noindent
(iii)  All the eigenvalues of $\hamG $ are of the form~\eqref{lan}, and $\Psi_\LB$ is eigenfunction of $\hamG $ with eigenvalue $E_n$ if and only if it is given by a linear combination of the functions~\eqref{psin} with $n_i\in\NN$ and $n_1+\cdots+ n_N=n$.
\end{theorem}

Therefore the bound states of this system satisfy
$$
 E_\infty= \lim_{n\to \infty}E_n=\frac{\om^2}{2\la}  ,\qquad  \lim_{n\to \infty}(E_{n+1}-E_n)=0.
 $$
Such a discrete spectrum is depicted in  figure~\ref{figure5} for several values of $\la$.

\begin{figure}
\includegraphics[height=7cm,width=10cm]{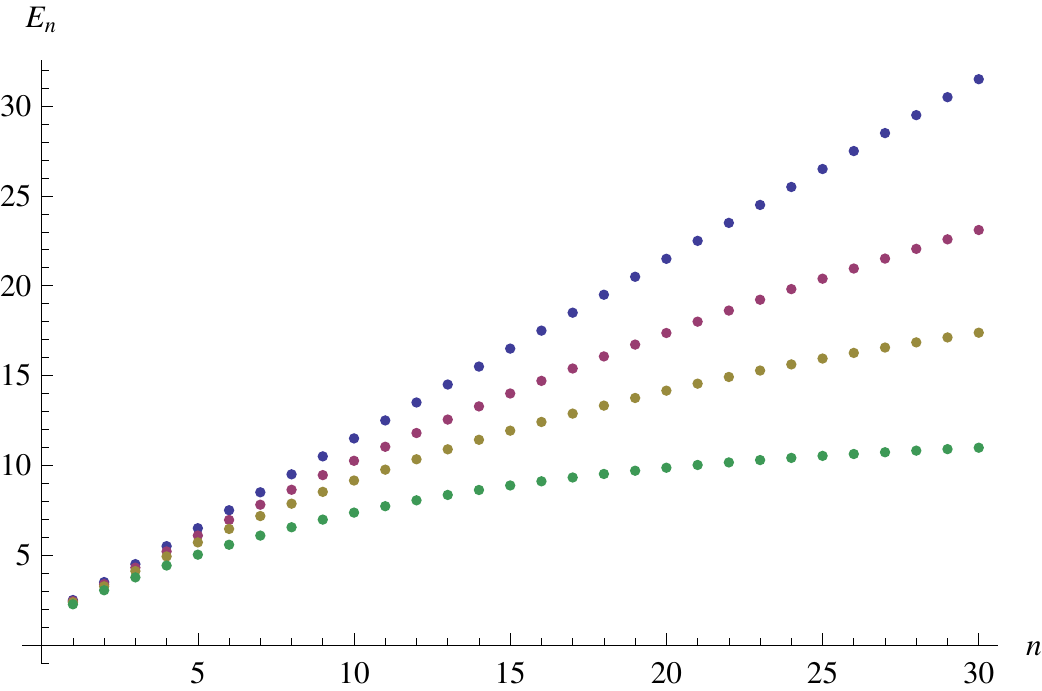}
\caption{The  discrete spectrum  (\ref{lan}) for   $0\le n\le 25$, $N=3$,  $\hbar=\om=1$ and $\la=\{0,0.01,0.02,0.04\}$ starting from the upper dot line   corresponding to the isotropic harmonic oscillator with $\la = 0$; in the same order, $E_0=\{ 1.5,  1.48, 1.46 , 1.41\}$ and $E_\infty=\{ \infty,50,25,12.5\}$.
 \label{figure5}}
\end{figure}


\sect{Concluding remarks}

Summarizing, we have presented a novel exactly solvable quantum nonlinear oscillator in $N$ dimensions, that can be understood as a simultaneous ``analytic" $\la$-deformation of both the usual isotropic oscillator potential and the underlying space on which the dynamics is defined. 
It turns out that if both sides of the Hamiltonian (the manifold and the potential) are appropriately modified, the curved quantum system preserves all the superintegrability properties of the Euclidean one, and its full solution can be explicitly obtained by making use of the curved analogues of Fradkin operators. It is worth stressing that such an explicit solution could be of interest from the physical viewpoint, since a parabolic effective-mass function has been proposed in~\cite{Koc,Schd} in order to describe realistic quantum wells formed by semiconductor heterostructures.

On the other hand, the quest for the preservation of superintegrability under quantization seems to be also a valuable guideline in order to clarify the properties of different possible quantization recipes on generic Riemannian manifolds. In this sense, we think that the connection here presented between the conformal Laplacian approach and the MS property   is worth to be investigated through, for instance, the study of the quantization of other MS systems on spaces of nonconstant curvature that have been recently characterized in the context of the generalized Beltrami theorem~\cite{commun}. 

Finally, we recall that the real parameter  $\la=1/\k$ was restricted in~\cite{PhysD}  to take a {\em positive} value.  However, the MS of the classical Hamiltonian stated in Theorem 1 does hold for {\em negative}  $\la$ as well.  Nevertheless, the underlying space and the oscillator potential change dramatically  when $\la<0$ (see~\cite{IJTP}), and the corresponding quantum problem is currently under investigation by making use of the techniques here presented.


\section*{Acknowledgments}

This work was partially supported by the Spanish MICINN   under grants    MTM2010-18556   and FIS2008-00209, by the   
 Junta de Castilla y
León  (project GR224), by the Banco Santander--UCM 
(grant GR58/08-910556)
   and by  the Italian--Spanish INFN--MICINN (project ACI2009-1083).


\end{document}